\documentclass[psfig]{aa}
\input psfig.tex

\begin{document}                                                                

\twocolumn


   \title{Non-Markov Excursion Set Model of Dark Matter Halo Abundances}

   \titlerunning{Non-Markov Excursion Set Model}

   \author{Grigori Amosov$^{(1)}$ and Peter Schuecker$^{(2)}$}
                                       
   \authorrunning{Amosov \& Schuecker}

   \offprints{Grigori Amosov: amosov@fizteh.ru, Peter Schuecker: peters@mpe.mpg.de}

   \institute{
    $^{(1)}$ Moscow Institute of Physics and Technology, Institutski
             9, 141700 Dolgoprudni, Russia\\
    $^{(2)}$ Max-Planck-Institut f\"ur extraterrestrische Physik,
             Giessenbachstra{\ss}e, 85741 Garching, Germany\\}

   \date{Received  ; accepted }                         
   
   \markboth{Non-Markov Excursion Set Model}{}

\abstract{The excursion set model provides a convenient theoretical
framework to derive dark matter halo abundances. This paper
generalizes the model by introducing a more realistic merging and
collapse process. A new parameter regulates the influence of the
environment and thus the coherence (non-Markovianity) of the merging
and the collapse of individual mass shells.  The model mass function
also includes the effects of an ellipsoidal collapse. Analytic
approximations of the halo mass function are derived for
scale-invariant power spectra with the slopes $n=0,-1,-2$. The $n=-2$
mass function can be compared with the results obtained from the
`Hubble volume' simulations. A significant detection of non-Markovian
effects is found for an assumed accuracy of the simulated mass
function of 10\%.  \keywords{galaxies: clusters: general -- cosmology:
theory -- dark matter} }

\maketitle

\section{Introduction}\label{INTRO}

The hierarchical growth of virialized cosmic structures provides a
useful physical paradigm for the understanding of the formation of
galaxies and clusters of galaxies in the Universe (White \& Rees
1978). The Zel'dovich (1970) theory also includes a description of
partially virialized structures like filaments and walls. In both
models, cosmic structures grow from initial Gaussian density
fluctuations via gravitational instability, leading after merging and
their (partial) virialization to mass functions which are used as
powerful statistical diagnostics.

Theoretical mass functions of dark matter halos are estimated from
N-body simulations with accuracies of 10-30\,\% (e.g. Jenkins et
al. 2001). More physical insights can be obtained from analytic
treatments based on Press-Schechter (PS) -like arguments (Press \&
Schechter 1974; Bond et al. 1991, hereafter BCEK). Alternative
derivations are based on, e.g., non-Gaussian statistics (Lucchin \&
Matarrrese 1988), or treat both fluctuations and interactions of
density perturbations within one process (Cavaliere \& Menci 1994), or
directly consider the non-linear regime (Valageas \& Schaeffer
1997). The inclusion of non-spherical dynamical approximations (Monaco
1995, Lee \& Shandarin 1998) and ellipsoidal collapse models (Sheth,
Mo \& Tormen 2001, hereafter SMT, 2002) modify the assumption of a
spherical collapse while preserving the simplicity of the original
excursion set idea.

\begin{figure}
\vspace{-1.7cm}
\centerline{\hspace{0.0cm}
\psfig{figure=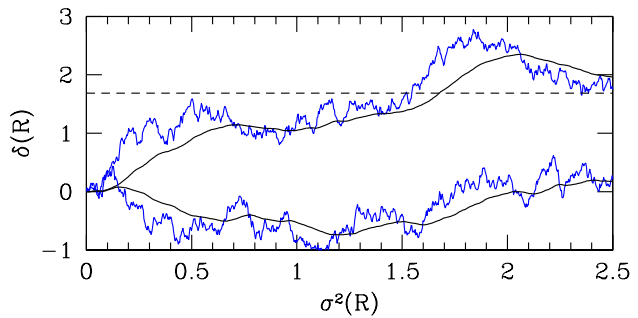,height=8.5cm,width=8.5cm}}
\vspace{-2.50cm}
\caption{\small Trajectories for the extended PS process (Markovian
process, jagged curves, $T=0$) and for the non-Markovian process
(smooth curves, $T=0.23$).}
\label{FIG_T}
\end{figure}

The excursion set model of BCEK assumes a Gaussian density field which
is smoothed at a given spatial location with a top-hat filter in
wavenumber space (sharp $k$-space filter) using different comoving
filter radii $R$. The resulting filtered density contrasts $\delta(R)$
perform a highly jagged diffusion trajectory (Fig.\,\ref{FIG_T}) where
the mass function is derived -- as a function of the standard
deviation $\sigma(R)$ of the mass density fluctuations -- from the
loss rate of trajectories at the barrier $\delta_{\rm c}$ at their
lowest $\sigma$ level (highest $R$ or mass scale). For the spherical
collapse model and for the Einstein-de Sitter Universe we have
$\delta_{\rm c}=1.686$, weakly dependent on cosmology. The halo mass
function is thus directly related to the first passage time
distribution of the trajectories.

There is a problem with this approach related to the assumption of a
sharp $k$-space filter. The filter gives a quite unrealistic mass
assignment scheme, with a growth of cosmic structure which depends
only on the mass of a halo it has within an infinitesimally small time
interval at a given cosmic epoch, and without any dependency on past
or future properties of the halo (Markov assumption). Therefore, the
merging events occur as completely uncorrelated, sudden, jumps in the
formation history. We will replace the sharp $k$-space filter by a
non-sharp filter. This leads to a more realistic mass assignment
scheme. The corresponding growth of cosmic structure depends on the
properties of the halo over a finite and future-directed time range
(non-Markov assumption).

The idea to invoke non-Markovian processes is not new. A discussion of
the physical consequences of the Markov assumption and why related
processes could fail can be found in White (1996, 1997). BCEK pointed
out the relation between the shape of a mass filter and the Markov
assumption. Their discussion of non-Markovian processes is, however,
mainly restricted to results obtained with Monte-Carlo experiments.

Using the traditional excursion set model as a guideline,
Sect.\,\ref{FIRST} introduces a simple analytic model which describes
a more uniform mass assignment scheme, and thus a more uniform
spherical collapse of dark matter halos. This model is further
improved in Sect.\,\ref{ELLIP}, by including the effects of an
ellipsodial collapse similar to SMT. The combined model can be
regarded as a simple though typical example of a non-Markovian
process. It generalizes the traditional excursion set model in a
manner such that non-Markovianity can now be gradually increased by a
new filter parameter. The same filter parameter also increases the
smoothness of the profile of the mass filter (see Eq.\,\ref{FILTER} in
Sect.\,\ref{FF}).

The standard mirror image method is used in Sect.\,\ref{MASSF} to
derive an analytic form for the halo mass function. We show why this
method, which in general does not work for non-Markovian processes,
can be used in our specific case. The resulting mass function has the
same functional form as the standard excursion set result in terms of
the variance of the mass distribution and the critical density
threshold. However, non-Markovian effects change the relations between
filter radius, mass and variance. These relations become a function of
the power spectrum of the underlying mass distribution. After the
corresponding transformations of the mass functions they also become
apparent in the halo mass function itself. This is the reason why in
our non-Markovian context, the profile of the mass filter and thus how
much mass is sweeped in from the surrounding mass of a collapsing
region becomes a function of the power spectrum.

How much mass the filter sweeps in for a given filter radius can be a
quite complex function, especially when general power spectra are
considered. We could derive approximate analytic results for
scale-invariant power spectra with the slopes $n=0,-1,-2$. The latter
case is close to the observed value and allows a comparison with the
Jenkins et al. (2001) mass function obtained from the `Hubble Volume'
simulations (Sect.\,\ref{DISCUSS}). The basic aim is to test under the
given assumptions the presence of non-Markovian effects in the
simulations and to determine how accurately mass functions should be
measured to detect the effects. A discussion of general power spectra
goes beyond the scope of our analytic treatment and is postponed to a
further paper.

\section{Spherical collapse and merger trajectories}\label{FIRST}

\begin{figure*}
\vspace{0.0cm}
\centerline{\hspace{-11.5cm}
\psfig{figure=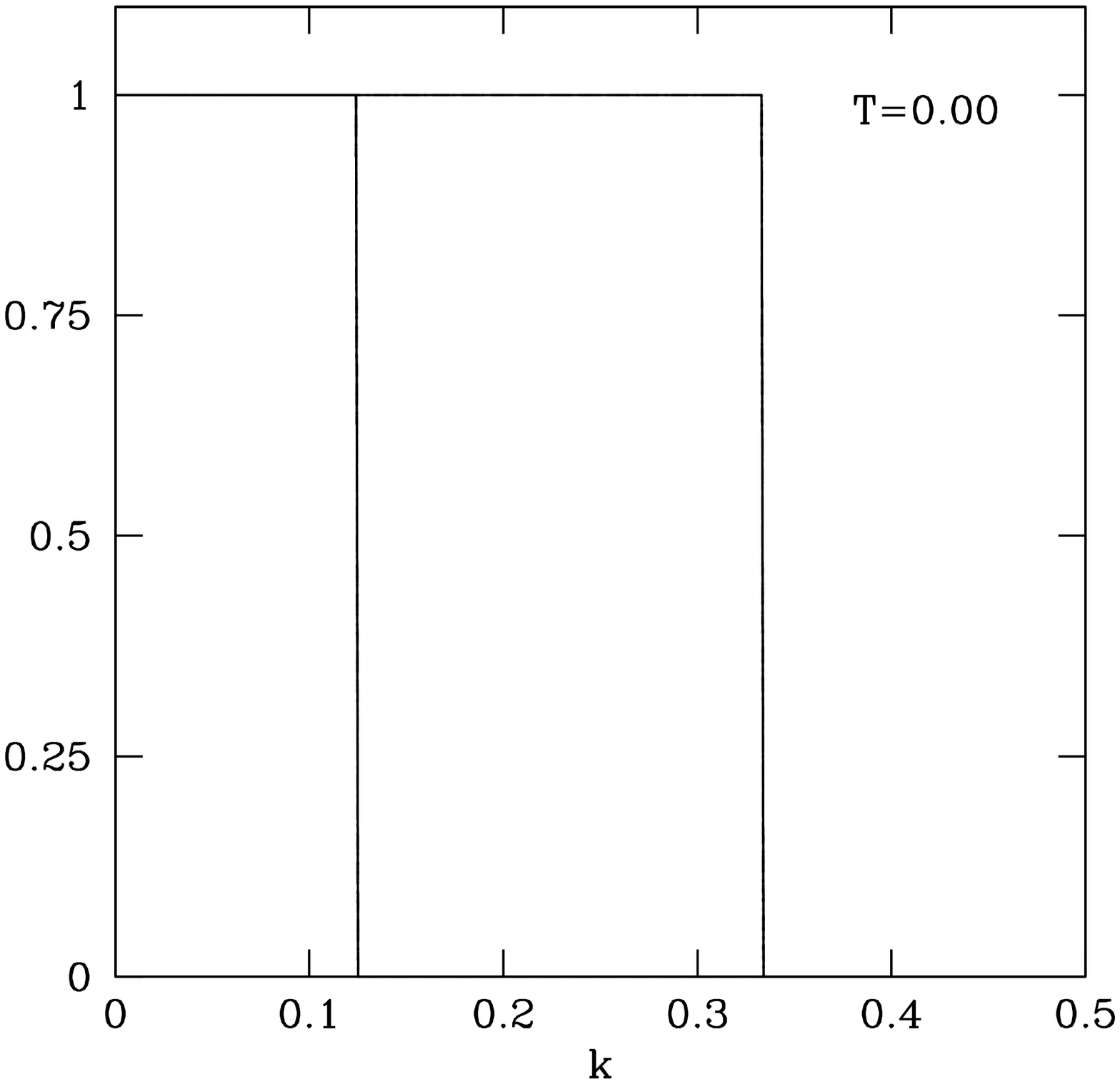,height=7.0cm,width=7.0cm}}
\vspace{-7.0cm}
\centerline{\hspace{0.25cm}
\psfig{figure=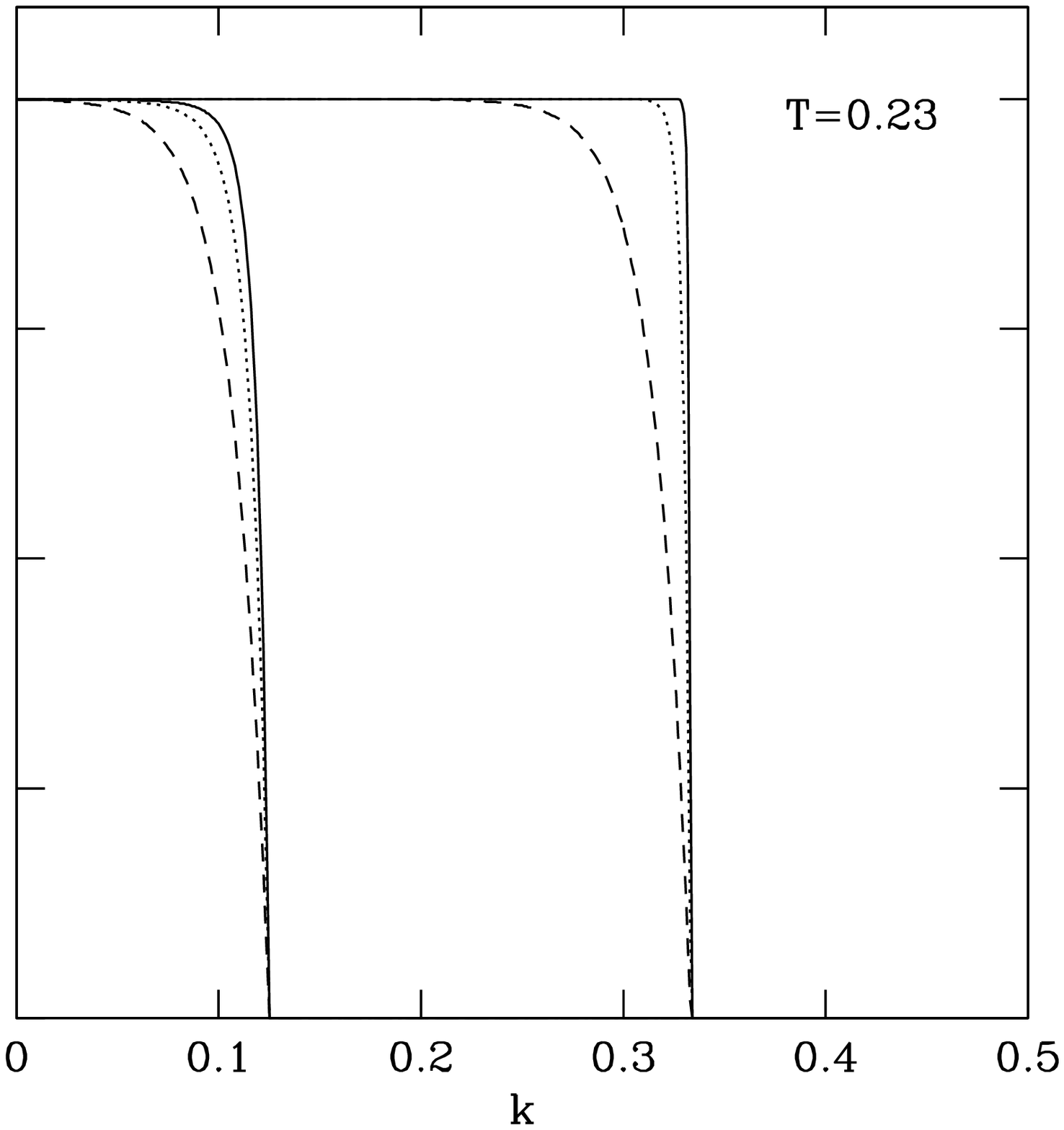,height=7.0cm,width=7.0cm}}
\vspace{-7.0cm}
\centerline{\hspace{11.5cm}
\psfig{figure=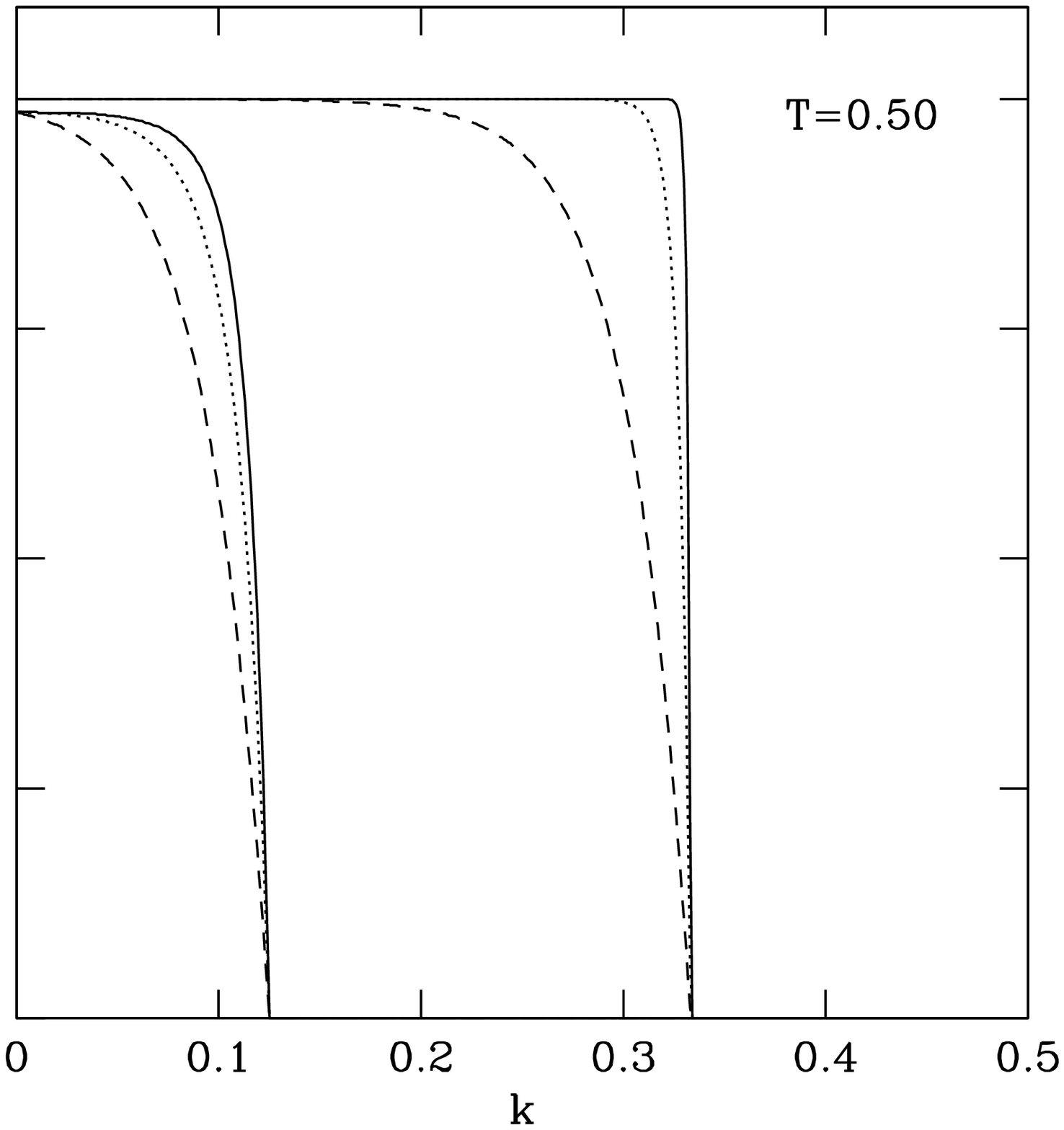,height=7.0cm,width=7.0cm}}
\vspace{-0.50cm}
\caption{\small Filter profiles $W_T(k)$ as a function of comoving
wavenumber $k$ in units of $[h\,{\rm Mpc}^{-1}]$ for three values of
the $T$ parameter (left: $T=0.00$, middle: $T=0.23$, right:
$T=0.50$). Each panel shows two groups of filters. Left group: filter
radius $R=8\,h^{-1}\,{\rm Mpc}$, right group: $R=3\,h^{-1}\,{\rm
Mpc}$. Each group consists of three filter curves. From up to down:
$n=0$ (continuous), $-1$ (dotted) ,$-2$ (dashed). The normalization of
the power spectrum $P_0$ is adjusted to give $\sigma_8=0.9$ (at
$T=0$).}
\label{FIG_F}
\end{figure*}

The Markov assumption corresponds to an oscillating filter in
configuration space or a sharp filter in $k$-space. For the
determination of the total collapsed mass of a halo, the formal
weighting of each mass shell with an oscillating filter suggests the
following collapse picture. The innermost mass shell located around a
peak in the cosmic mass distribution (though not necessarily around a
massive particle) is up-weighted in the mass budget and is thus
expected to collapse, whereas the next overlaying mass shell located
around the first minimum of the filter is down-weighted, the following
mass shell is up-weighted again, and so on. Similarily, for the
determination of merger trajectories, the sharpness of the filter in
$k$-space leads to the addition of statistically independent
fluctuation power in disjunct $k$-shells and thus to highly jagged
trajectories with completely independent increments. Both the collapse
and the merger trajectories are regard as unrealistic
(Fig.\,\ref{FIG_T}).

BCEK showed that a smooth, non-oscillating filter can be obtained with
a non-Markovian process. It is also well-known that smooth filters
like a Gaussian or a standard top-hat are consistent with the uniform
(coherent) collapse of all concentric mass shells. Moreover, the
smoothness of the filter in $k$-space mixes fluctuation power located
in disjunct $k$-shells and thus smoothes the merger trajectories. We
are thus searching for diffusion processes with smooth merger
trajectories. The physical significance of the corresponding mass
filter is evaluated by the comparison with mass functions derived from
simulations (Sect.\,\ref{DISCUSS}).

Formally, the merger trajectories of the excursion set model are
described by the Wiener stochastic diffusion process $W(t)$ -- the
Brownian motion. We follow the general convention and work with a
pseudo time or mass resolution variable $t(R)$ in $\sigma(t)$ which
will be specified later (Eq.\,\ref{RESOL}), and the filtered density
contrast $\delta(\sigma)$ (see BCEK and Lacey \& Coles 1993 for more
details). A non-Markovian process with the desired properties has the
increments of the density contrast
\begin{eqnarray}\label{MEM}
d\delta (t)=\left[ \frac {1}{T}\int \limits _0^te^{-\frac
{t-s}{T}}dW(s) \right]dt\,.
\end{eqnarray}
The increments depend on all values of $W(t)$ within $[0,t]$, and if
the parameter $T$ tends to zero, the smoothness of the merger
trajectories deminishes (Fig.\,\ref{FIG_T}).  Whereas the sum over
history is in most cases a generic feature of non-Markovian processes,
the kernel function in Eq.\,(\ref{MEM}) may not necessarily be
exponential. However, its explicit form comes from the shape of the
mass filter function (Eq.\,\ref{FILTER}) and finds its justification
through the usuability of this mass filter. Its shape suggests more
realistic extraction of the material which will ultimately end up in a
spherical collapse to form a virialized halo. The new parameter $T$
measures the coherence (non-Markovianity) of the collapsing mass
shells and can be estimated from fits of halo mass functions to
simulated or observed data (Sect.\,\ref{DISCUSS}).

\section{Ellipsoidal collapse and non-Markovianity}\label{ELLIP}

The theory of ellipsoidal collapse (Bond \& Myers 1996, SMT) provides
a relation between the critical density contrast of an ellipsoidal
collapse $\delta_{\rm ec}$ and a spherical collapse $\delta_{\rm c}$
of the form $\delta_{\rm ec}=\delta_{\rm c}+\beta\delta_{\rm
c}(\sigma/\delta_{\rm c})^{2\gamma}$. For convenience we set
$\gamma=1$, yielding mathematically exact first passage time
distributions. Our approximation
\begin{equation}\label{T6}
\delta_{\rm ec}\,=\,\delta_{\rm c}\,+\,\frac{\beta}{\delta_{\rm
c}}\,\sigma^2
\end{equation}
deviates from the exact result on the 10\,\% level or better for
$\beta=0.3$ within $0.0<\sigma^2<1.5$. In the diffusion model of
merger trajectories, the ellipsoidal collapse (Eq.\,\ref{T6}) induces
a linear drift superposed onto the diffusion process (Eq.\,\ref{MEM})
where high (low) mass objects needs low (high) effective critical
density contrasts for their virialization. We thus rewrite
Eq.\,(\ref{MEM}) as
\begin{eqnarray}\label{F1}
d\delta (t)=\left[ \frac {1}{T}\int \limits _0^te^{-\frac {t-s}{T}}dW(s)
\right]dt-\beta\,d\sigma^2(t)\phantom{XXXX}\nonumber \\
=\frac{1}{T}\,\left\{\,W(t)\,-\,\left[\delta(t)-\rm
E(\delta (t))\right]\,\right\}dt-\beta\,\,\frac{d}{dt}{\rm
Var}(\delta(t))\,dt\,,\nonumber\\ \delta (0)=0\,,\quad
\end{eqnarray}
where $T$ and $\beta$ are fixed positive numbers. The solution of
Eq.\,$(\ref{F1})$ can be represented as
\begin{equation}\label{F2}
\delta (t)=\int \limits _{0}^t(1-e^{-\frac
{t-s}{T}})\,dW(s)-\beta\sigma ^2(t)\,,
\end{equation}
with the expectation ${\rm E}(\delta (t))\,=\,-\beta\sigma ^2(t)$ and
the variance
\begin {eqnarray}\label{F25}
\sigma ^2(t)={\rm
Var}(\delta (t))=\int \limits _0^t(1-e^{-\frac
{t-s}{T}})^2\,ds\,\nonumber\\ =t-\frac {3}{2}T+2Te^{-\frac
{t}{T}}-\frac {T}{2}e^{-\frac {2t}{T}}\,.
\end {eqnarray}

For some applications (Eqs.\,\ref{VOL5},\ref{VOL7},\ref{VOL3}) it is
convenient to have the inverse of Eq.\,(\ref{F25}). A good
approximation is
\begin{eqnarray}\label{INV}
t(\sigma^2)\,=\,\sigma^2(t)\,+\,1.98\,T\, \left\{\,1\,-\,
e^{-\left[\frac{\sigma^2(t)}{T}\right]^{0.363}}\,\right\}\,,
\end{eqnarray}
which has an accuracy of better than 2\% for $\sigma\le 1$ and $0\le T
<5$.

The smoothness (differentiability) of the sample trajectories
(Fig.\,\ref{FIG_T}) is a clear signature of the non-Markovianity of
the process (Eq.\,\ref{F2}). A general discussion of non-Markovianity
can be found in Feller (1968). As mentioned above, a non-Markovian
process can be obtained by integrating certain Markov processes. It is
well-known from diffusion theory that if one starts with the Markov
process $V(t)$ representing particle velocities in a viscid medium,
then positions $X(t)$ defined as $dX(t)=V(t)dt$ form the simplest
non-Markovian process. Note that setting a complete path $X(s),\
t-T\leq s\leq t,$ is not the same as setting only one position $X(t)$
which provides the non-Markov property of $X(t)$. In our case, the
process $\delta(t)$ can be represented as an integral over the
Ornstein-Uhlenbeck process (see Sect.\,\ref{MASSF}) and is thus
non-Markovian. Moreover, the $\delta (t)$ are Gaussian and have a
probability distribution which is similar to the probability
distribution of $W(t)-\beta t$.

\section{Filter function}\label{FF}

The mass filter of the non-Markovian process can be derived as in
Schuecker et al. (2001a) by equating the process variance
(Eq.\,\ref{F25}) and the variance of the underlying matter
fluctuations,
\begin{equation}\label{EQ}
\int_0^t\,ds\,\left(1-e^{\frac{s-t}{T}}\right)^2\,=\,
\int_0^\infty\frac{4\pi\,k^2\,dk\,P(k)}{(2\pi)^3}\,\left|W_T(kR)\right|^2\,,
\end{equation}
where $P(k)$ is the matter power spectrum, $W_T(kR)$ the mass filter
function in $k$-space, and $R$ the filter scale in configuration
space. Eq.\,(\ref{EQ}) cannot be solved easily for an arbitrary
$P(k)$. However, for scale-invariant power spectra of the form
$P(k)=P_0\,k^n$ the filter
\begin{equation}\label{FILTER}
W_T(kR)=1-\exp{\left\{\frac{P_0}{2\pi^2(n+3)R^{n+3}}
\left[\frac{(kR)^{n+3}-1}{T}\right]\right\}}\,,
\end{equation}
for $kR\le 1$, and $W_T(kR)=0$ elsewhere, is consistent with
Eq.\,(\ref{F25}) when the resolution variable is defined as
\begin{equation}\label{RESOL}
t\,=\,\frac{P_0}{2\pi^2(n+3)R^{n+3}}\,.
\end{equation}
Here, $t$ is the same as in the Markov excursion set model and
requires $n>-3$ for structure to grow through hierarchical clustering.

A new aspect is the dependence (for $T>0$) of the filter on the power
spectrum (Fig.\,\ref{FIG_F}). For standard filters, the profile does
not change with $P(k)$ and only normalization factors like the
characteristic mass-scale $M_*$ used to get convenient invariance
properties of the halo mass function depend on $P(k)$. In the present
case the filter profile depends on $P(k)$ and thus on the spatial
correlation function.  The correlation function approximates the mean
shape of density peaks. Therefore, the profile of the filter roughly
follows the mean profile of density peaks.

\subsection{Filter volumes}

From the volume $V_T$ of the filter we can estimate the mass
$M=\bar{\rho}V_T$ extracted by the filter from the mean cosmic matter
density $\bar{\rho}$.  The volume $V_T(R)$ is related to the Fourier
transform of Eq.\,(\ref{FILTER}) by

\begin{figure*}
\vspace{0.0cm}
\centerline{\hspace{-11.5cm}
\psfig{figure=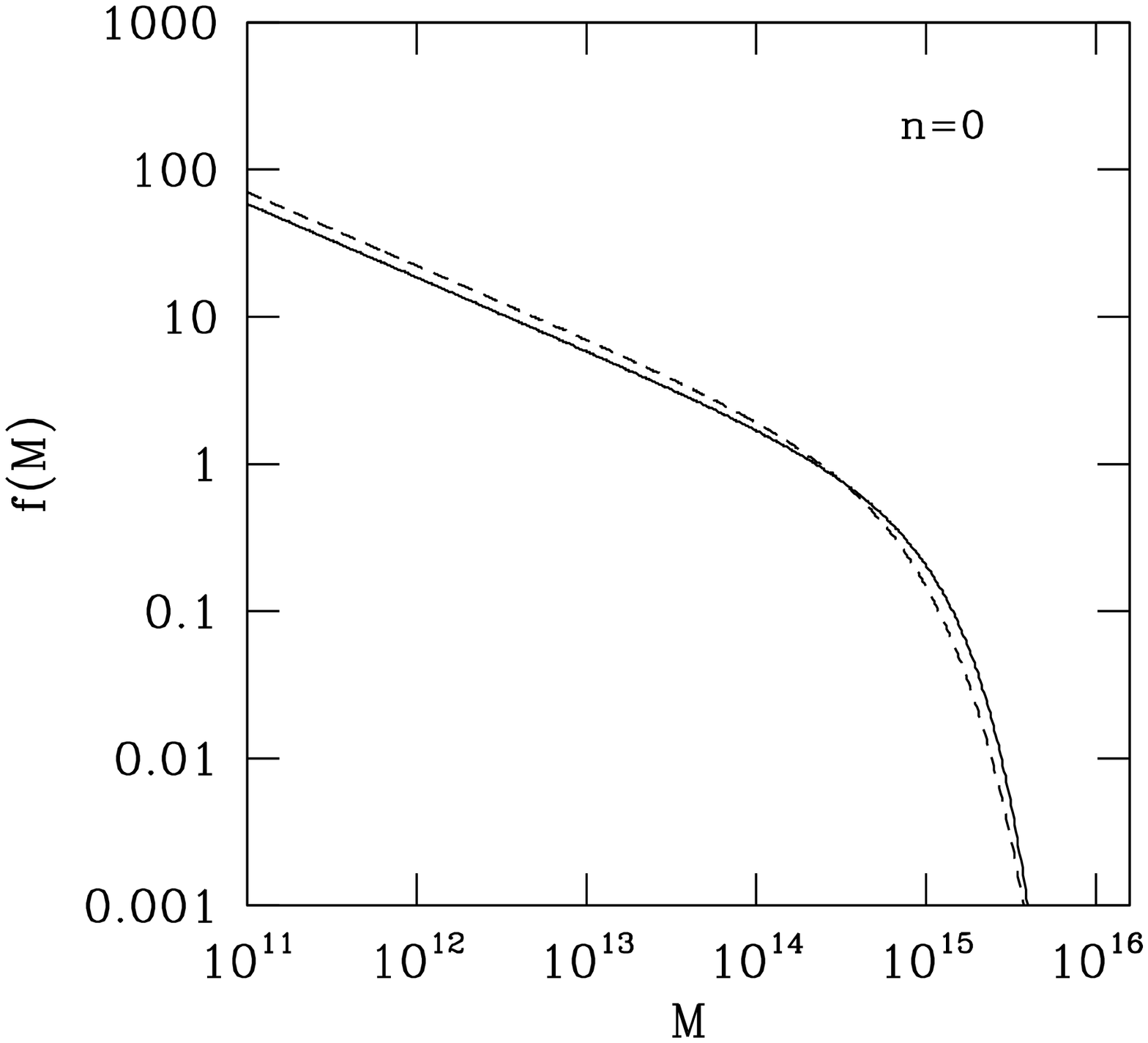,height=7.0cm,width=7.0cm}}
\vspace{-7.0cm}
\centerline{\hspace{0.25cm}
\psfig{figure=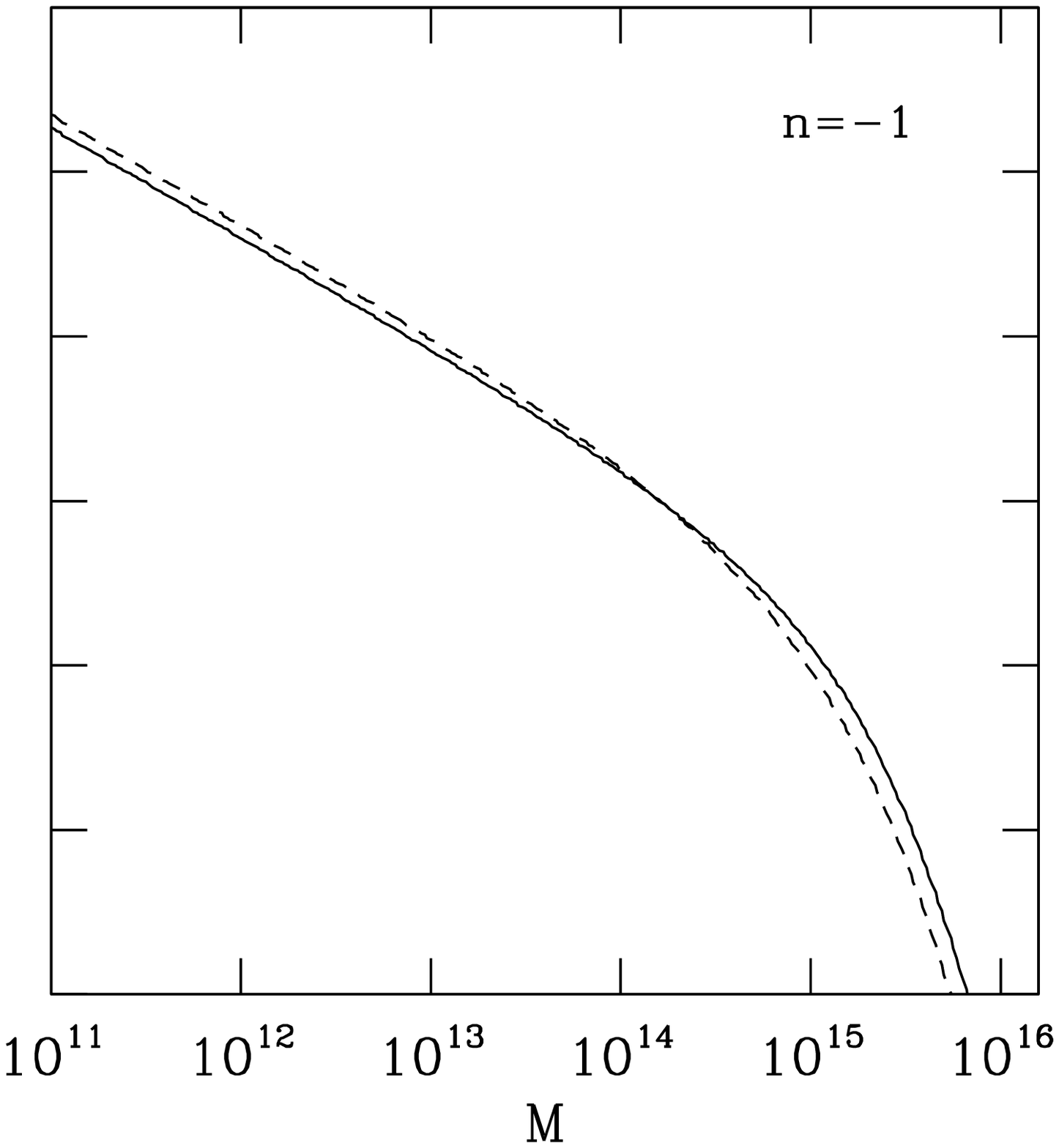,height=7.0cm,width=7.0cm}}
\vspace{-7.0cm}
\centerline{\hspace{11.5cm}
\psfig{figure=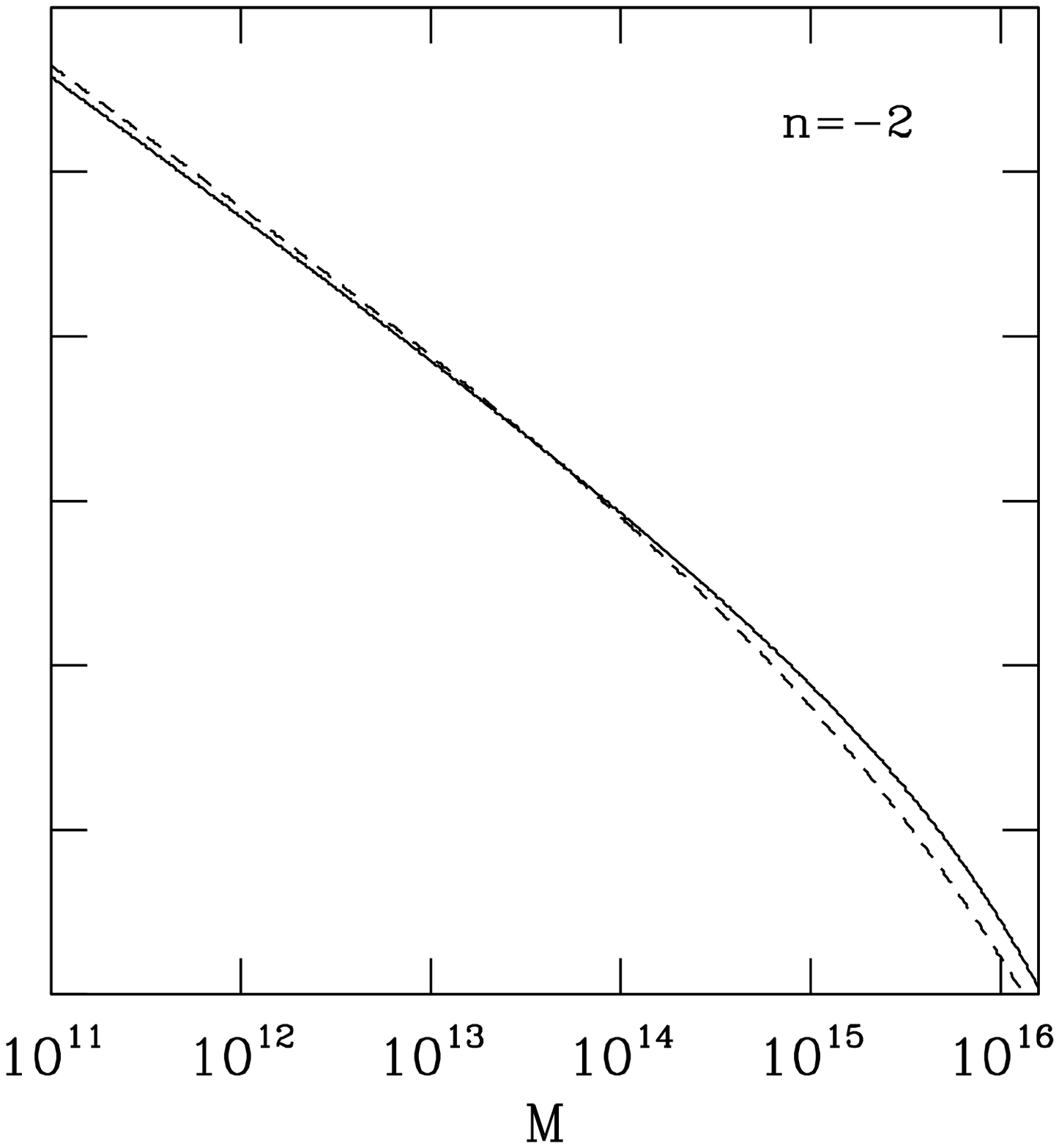,height=7.0cm,width=7.0cm}}
\vspace{-0.75cm}
\caption{\small Mass functions $f(M)$ for Gaussian fluctuation fields
with a power spectrum with $n=0,-1,-2$. Mass $M$ in solar units:
$T=0$, $\beta=0$ (dashed lines), and $T=0.23$, $\beta=0$ (continuous
lines). The computations assume the normalized cosmic mass density
$\Omega_m=0.27$, the standard normalization $\sigma_8=0.9$ and the
normalized Hubble constant $h=0.7$.}
\label{FIG_MF}
\end{figure*}

\begin{equation}\label{VOL1}
\frac{1}{V_T(R)}=W_T(r=0,R)=\frac{1}{2\pi^2}
\int_0^{1/R}dk k^2 W_T(kR)\,.
\end{equation}
For the filter (Eq.\,\ref{FILTER}), three cases can be studied in a
straightforward manner. For a power spectrum with the spectral index
$n=0$ we have
\begin{equation}\label{VOL4}
\frac{1}{V_T}=\frac{1}{6\pi^2R^3}-\frac{T}{P_0}\left(1-
e^{-\frac{P_0}{6\pi^2TR^3}}\right)\,,
\end{equation}
\begin{equation}\label{VOL5}
P_0=3072\pi^2\left[\sigma_8^2+1.98\,T\left(1-
e^{-(\frac{\sigma_8^2}{T})^{0.363}}\right)\right]\,{\rm Mpc}^3\,.
\end{equation}
For $n=-1$ integration of Eq.\,(\ref{VOL1}) yields
\begin{eqnarray}\label{VOL6}
\frac{1}{V_T}=\frac{1}{6\pi^2R^3}-\frac{T}{P_0 R}+
\left(\frac{\pi T}{P_0}\right)^{3/2}\,\times\nonumber\\e^{-\frac{P_0}{4\pi^2R^2T}}\,\,
{\rm erfi}\left(\frac{1}{2\pi R}\sqrt{\frac{P_0}{T}}\right)\,,
\end{eqnarray}
\begin{equation}\label{VOL7}
P_0=256\pi^2\left[\sigma_8^2+1.98\,T\left(1-
e^{-(\frac{\sigma_8^2}{T})^{0.363}}\right)\right]\,{\rm Mpc}^2\,.
\end{equation}
For $n=-2$ integration of Eq.\,(\ref{VOL1}) yields
\begin{equation}\label{VOL2}
\frac{1}{V_T}=\frac{1}{6\pi^2R^3}-\frac{T}{P_0R^2}+\frac{4\pi^2T^2}{P_0^2R}
-\frac{8\pi^4T^3}{P_0^3}\left(1-e^{-\frac{P_0}{2\pi^2RT}}\right)\,,
\end{equation}
\begin{equation}\label{VOL3}
P_0=16\pi^2\left[\sigma_8^2+1.98\,T\left(1-
e^{-(\frac{\sigma_8^2}{T})^{0.363}}\right)\right]\,{\rm Mpc}\,,
\end{equation}
where Eqs.\,(\ref{VOL5},\ref{VOL7},\ref{VOL3}) use the approximation
(Eq.\,\ref{INV}), and $\sigma_8$ is a parameter determined by the
filter (Eq.\,\ref{FILTER}) analogous to the standard normalization of
the power spectrum. In Eq.\,(\ref{VOL6}), erfi is the imaginary error
function. In all three cases, for a given radius the volume and thus
the mass covered by a filter with $T>0$ is larger compared to $T=0$
where we have $V_0=6\pi^2R^3$.

\subsection{Filter radius-mass relations}

For the mass function, the relation between the filter radius and mass
are needed. Therefore, Eqs.\,\,(\ref{VOL4},\ref{VOL6},\ref{VOL2}) have
to be solved for $R$. For $n=0$ we have
\begin{eqnarray}\label{RM2}
R(M)&=&\frac{M^{1/3}}{\left\{6\pi^2\bar{\rho} +
\frac{6\pi^2TM}{P_0}\left[1+W(z)\right]\right\}^{1/3}}\,,\\
z&=&-\frac{1}{\exp\left(1+\frac{\bar{\rho}P_0}{MT}\right)}\,,\nonumber
\end{eqnarray}
with $W(z)$ the principal branch of the Lambert W-function
(e.g. Corless et al. 1996, not to be confused with the Wiener process
in Eq.\,\ref{F1}). In the present case we have $\infty> T \ge 0$ which
corresponds to the range $-1/e< z\le 0$ where $W(z)$ is monotonically
increasing with values in the interval $-1< W(z)\le 0$. In some cases
(Eq.\,\ref{F9}) the derivative $dW(z)/dz=W/[z(1+W)]$ is needed.

For $n=-1$ we can neglect the final exponential term in
Eq.\,(\ref{VOL6}) which has an effect of $<5\%$ for $T\le 1$ up to
halo masses of $10^{16}\,M_\odot$. Solving the truncated equation for
$R$ yields
\begin{equation}\label{RM3}
R(M)=\frac{(U_1+3\sqrt{U_2})^{1/3}}{U_3}\,+\,
\frac{U_4}{(U_1+3\sqrt{U_2})^{1/3}}\,-\,U_5\,,
\end{equation}
where the functions $U_1\ldots U_5$ are given in Appendix \ref{APP}
(Eqs.\,\ref{UU} etc.).

For the case $n=-2$ the final exponential term in Eq.\,(\ref{VOL2})
can be neglected which has an effect of $<5\%$ for $T\le 5$ up to halo
masses of $10^{16}\,M_\odot$. Note that Eq.\,(\ref{VOL2}) without the
exponential term can also be obtained by asymptotic expansion of
Eq.\,(\ref{VOL1}), which is important for generalizing the present
discussion to other power spectra. Solving the truncated equation for
$R$ yields the radius-mass relation
\begin{equation}\label{RM1}
R(M)=\frac{S_1\,S_2^{1/3}\,+\,S_3\,-\,S_4\,S_2^{-1/3}}{S_5}\,,
\end{equation}
where the functions $S_1\ldots S_5$ are given in Appendix \ref{APP}
(Eqs.\,\ref{SS} etc.).

\begin{figure}
\vspace{-0.5cm}
\centerline{\hspace{0.0cm}
\psfig{figure=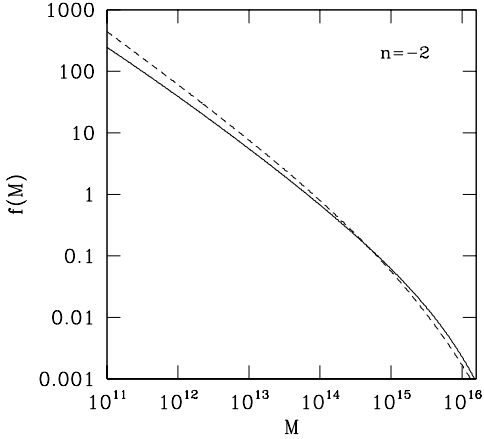,height=7.0cm,width=7.0cm}}
\vspace{-0.75cm}
\caption{\small Mass functions $f(M)$ for Gaussian fluctuation fields
with a power spectrum with $n=-2.0$. Mass $M$ in solar units. $T=0$,
$\beta=0$ (dashed line), and $T=0.23$, $\beta=0.12$ (continuous
line). Further normalizations as in Fig.\,\ref{FIG_MF}.}
\label{FIG_MF3}
\end{figure}

\section{Halo mass function}\label{MASSF}

We define the halo mass function $n(M)=f(M)\bar{\rho}/M$ as the number
of dark matter halos per unit volume and mass. As first shown in BCEK,
the fraction $f(M)$ may be obtained from the first passage time
distribution function of the assumed diffusion process.

Let $\tau $ be a random variable which determines the first passage
time when $\delta (t)$ achieves the critical value $\delta _{\rm c}$.
Taking into account that $\delta (t)\in {\cal N}(-\beta \sigma
^2(t),\sigma ^2(t))$, let us define the process $\Delta (t)$ obtained
by replacing the origin to $2\delta _c$ such that $\Delta (t)\in {\cal
N} (2\delta _c-\beta \sigma ^2(t),\sigma ^2(t))$. Both processes are
Gaussian.  Calculating the probabilities $p(t)$ and $P(t)$ of the
trajectories of $\delta (t)$ and $\Delta (t)$ which attain the level
$\delta _c$ at the same moment $t$, we obtain
\begin {equation}\label{S_4}
p(t)\,=\,P(t)\,\exp(-2\delta _c\beta)\,.
\end {equation}
The set of trajectories of $\delta (s),\,0\le s\le t,$ with asymptotic
end points $\delta (t)\in (-\infty ,\delta _{\rm c})$ which attain the
level $\delta _{\rm c}$ at the time $\tau,\,\tau <t$ can be associated
with the set of trajectories of $\Delta(s),\,0\le s\le t$, which have
the asymptotic end points $\Delta (t)\in (-\infty ,\delta _{\rm
c})$. In fact, by means of the inclusion $\Delta (t)\in (-\infty
,\delta _c)$ it follows that there exists the time $\tau,\,\tau <t,$
such that $\Delta (t)=\delta _{\rm c}$. The Eq.\,(\ref{F1}) implies
that $d\delta (t)=V(t)dt-\beta d\sigma ^2(t)$, where $V(t)=\frac
{1}{T}\int \limits _0^te^{-\frac {t-s} {T}}dW(s)$ is the
Ornstein--Uhlenbeck process. Hence, increments of the processes
$\delta (t)$ and $\Delta(t)$ are driven by the Markov process
$V(t)$. This means that given a trajectory of $\delta(t)$, which goes
through the point $\delta _c$ at the moment $t=\tau$, there exists the
trajectory of $\Delta(t)$ which coincides with the trajectory of
$\delta(t)$ for $t\geq \tau $.  The Markov property of $V(t)$
guarantees a possibility to construct this trajectory.  It is thus
sufficient to give $\Delta(t)$ the same values of the driving
Ornstein--Uhlenbeck process $V(t)$ as $\delta(t)$ for $t\geq \tau
$. On the other hand, the probability of trajectories $\delta (t)$ and
$\Delta (t)$ which achieve the level $\delta _{\rm c}$ at the same
moment $\tau $ are connected by the Eq.\,(\ref{S_4}). Hence, the
relation (\ref{S_4}) implies
\begin{eqnarray}\label{S_5}
P(\delta(t)\le\delta _{\rm c}\,|\,\delta(\tau)=\delta_{\rm
c},\,\tau<t)=e^{-2\delta _{\rm
c}\beta} P(\Delta(t)\le\delta _c)\nonumber\,,
\end{eqnarray}
so that 
\begin{eqnarray}
F(t)=P(\tau \geq t)=P(\delta (t)\leq \delta _{\rm c})-
P(\delta(t)\le\delta_{\rm c}\,|\,\tau<t)=\nonumber\\ \frac{1}{\sqrt
{2\pi \sigma^2(t)}} \left\{\int \limits _{-\infty }^{\delta_{\rm c}}
\exp{\left[-\frac{(x+\beta\sigma ^2(t))^2}{2\sigma^2(t)}\right]}
dx\right.-\nonumber\\ \left.\exp(-2\delta _{\rm c}\beta)\int \limits
_{-\infty }^{\delta _{\rm c}}\exp{\left[-\frac {(x-2\delta _{\rm
c}+\beta\sigma ^2(t))^2}{2\sigma ^2(t)}\right]}dx\, \right\} \nonumber\,.
\end{eqnarray}
For the normalized loss rate of trajectories,
$f(t)=-\frac{dF(t)}{dt}$, we obtain the density
\begin{eqnarray}\label{F7}
f(t)dt\,&=&\,\frac{1}{\sqrt{2\pi}}\,\frac{\delta_{\rm c}}{\sigma^3(t)}\,
\exp{\left \{-\frac{[\delta_{\rm c}+\beta\sigma
^2(t)]^2}{2\sigma^2(t)}\right \}}\,d\sigma^2\,,\\
        &=&\,f(\sigma^2)\,d\sigma^2\,.\nonumber
\end{eqnarray}
Written in this form, the theoretical mass function shows no
dependency on $T$ and reproduces for $\beta=0$ the prediction of the
Markovian excursion set model. However, $\sigma^2(t)$, which is
basically a function of mass, depends on $T$ (see Eq.\,\ref{F25}) so
that the non-Markovianity becomes apparent after the transformation of
Eq.\,(\ref{F7}) into
\begin{equation}\label{F8}
f(M)\,=\,f(\sigma^2(M))\,\left|\frac{d\sigma^2(M)}{dM}\right|\,.
\end{equation}
Unfortunately, the final formulae of the form
\begin{equation}\label{F9}
f(M)\,=\,f(\sigma^2(R(M)))\,
\left|\frac{d\sigma^2(R)}{dR}\right|\,
\left|\frac{dR(M)}{dM}\right|\,
\end{equation}
become quite clumsy after the insertion of all relevant equations (see
Appendix \ref{APP}).

\begin{figure}
\vspace{-0.5cm}
\centerline{\hspace{0.0cm}
\psfig{figure=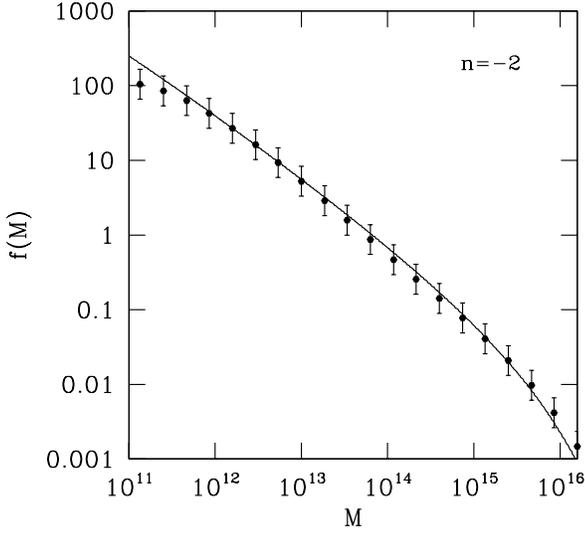,height=8.5cm,width=8.5cm}}
\vspace{-1.0cm}
\caption{\small Best fit mass function $f(M)$ with $T=0.23$,
$\beta=0.12$ (continuous line) and the transformed mass function of
Jenkins et al. (2001) with 20\% Gaussian random errors (dots with
error bars). Normalizations as in Fig.\,\ref{FIG_MF}.}
\label{FIG_M2}
\end{figure}

\begin{figure*}
\vspace{0.0cm}
\centerline{\hspace{-11.5cm}
\psfig{figure=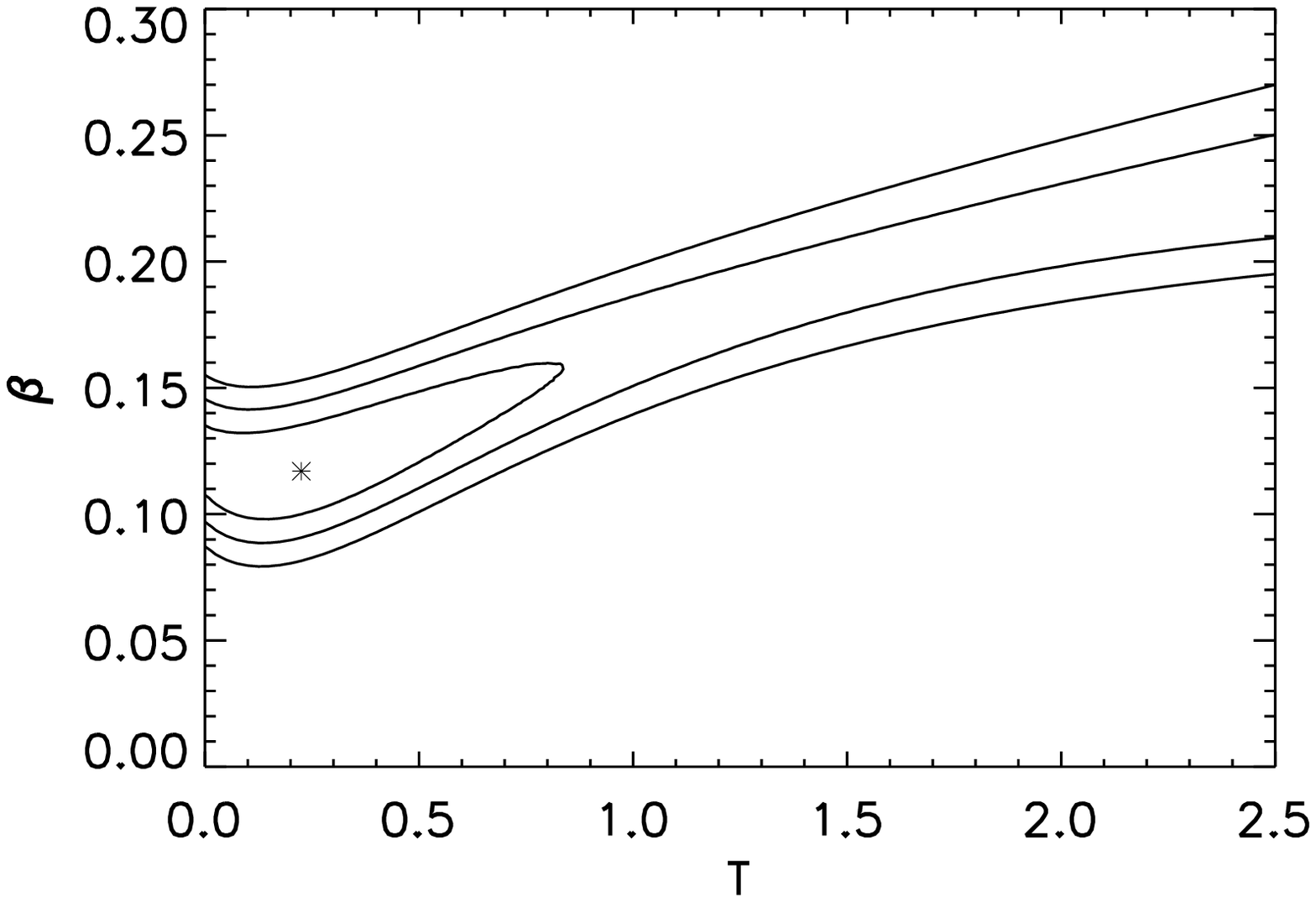,height=6.0cm,width=6.0cm}}
\vspace{-6.0cm}
\centerline{\hspace{0.25cm}
\psfig{figure=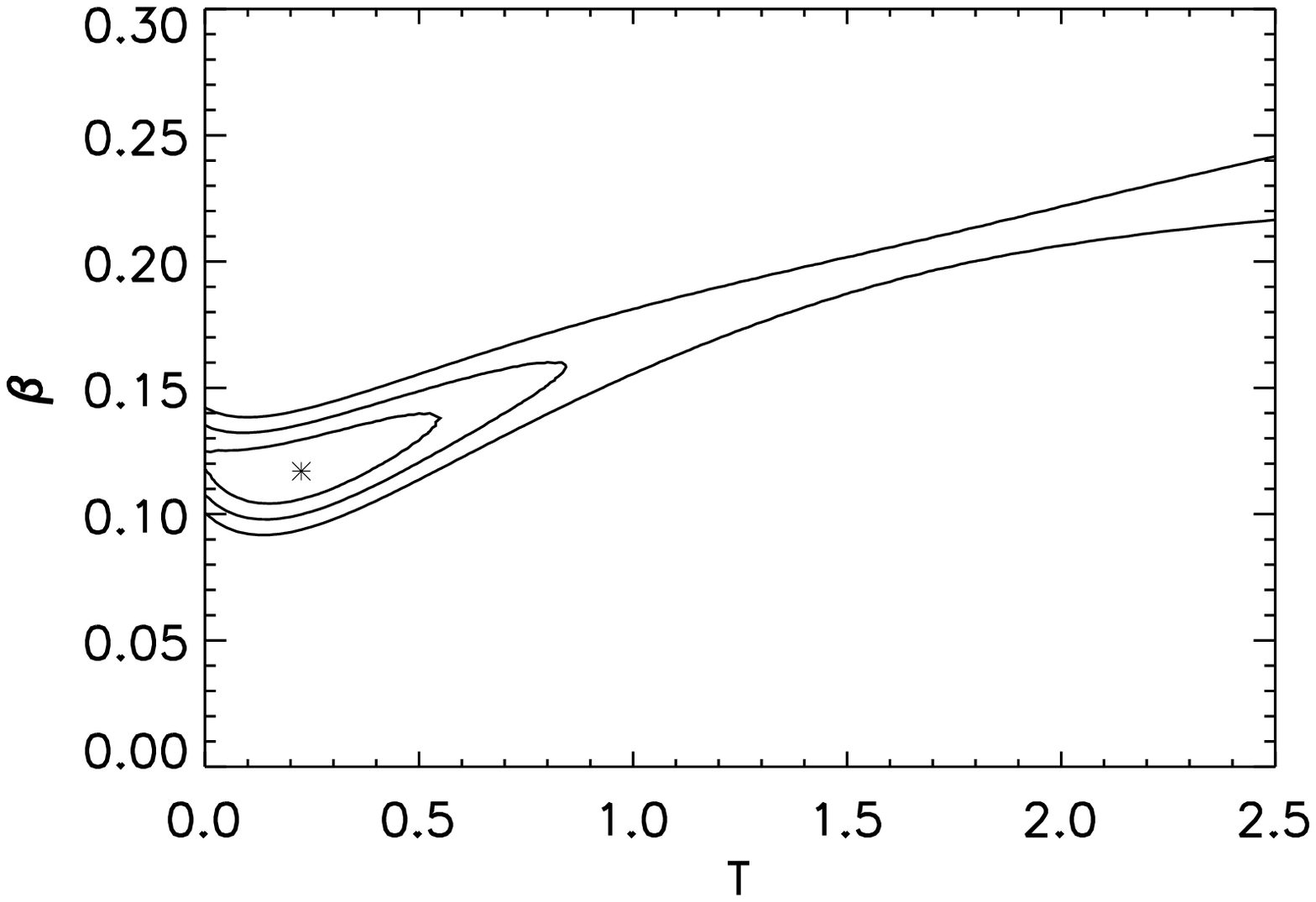,height=6.0cm,width=6.0cm}}
\vspace{-6.0cm}
\centerline{\hspace{11.5cm}
\psfig{figure=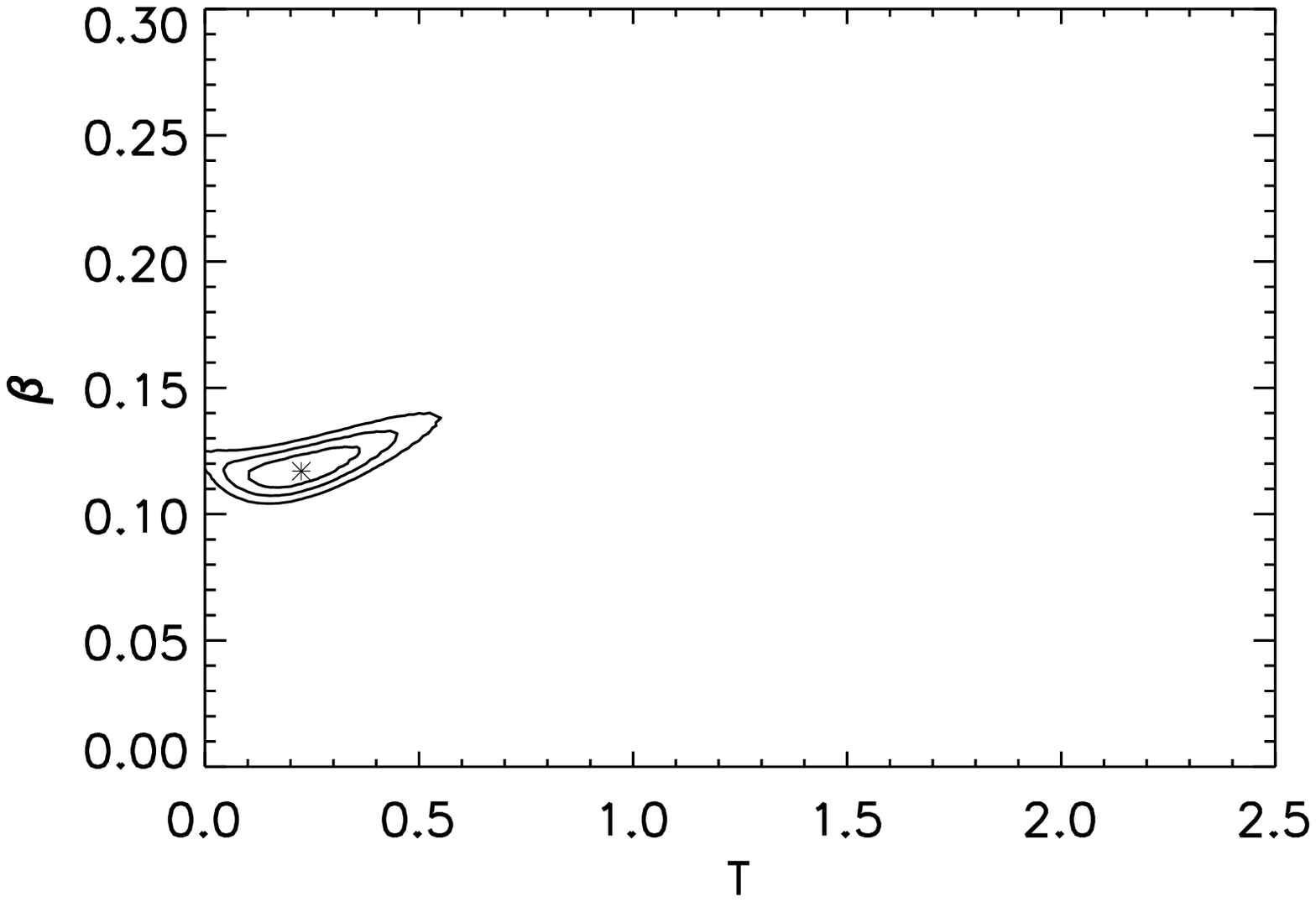,height=6.0cm,width=6.0cm}}
\vspace{-0.30cm}
\caption{\small $\chi^2$ distribution ($1-3\sigma$ contours for three
parameters) for the spectral index $n=-2$ obtained from the comparison
of mass functions $f(M)$ with different $T$ and $\beta$ parameter
values and the transformed Jenkins et al. (2001) mass
function. Normalizations as in Fig.\,\ref{FIG_MF}. The $\chi^2$ test
assumes errors of 30\% (left), 20\% (middle), 10\% (right) of the
Jenkins et al. mass function.}
\label{FIG_X2}
\end{figure*}

\section{Discussion of results}\label{DISCUSS}

The process (Eq.\,\ref{F2}) corresponds to a smooth filter profile and
allows a simple discussion of non-Markovian effects within the
excursion set model. The model includes the standard excursion set
result as a limiting case, and allows non-Markovian effects to be
increased gradually by the new filter parameter $T$.

\subsection{Effects of non-Markovianity}

Figure \ref{FIG_MF} shows halo mass functions $f(M)$ for different
power spectra and values of the $T$ parameter. In all cases the
non-Markovian effects described by Eq.\,(\ref{F2}) increase (decrease)
the number density of high (low) mass halos. At high masses the
ellipsoidality slightly decreases the number density of high mass
halos and thus partially compensates non-Markovianity. At small masses
both non-Markovianity and ellipsoidality decrease the number density
(Fig.\,\ref{FIG_MF3}).

The result of equating the mass and the process variances in
Eq.\,(\ref{EQ}) is a mass filter with a profile which depends on the
power spectrum of the mass distribution and thus on the mean profile
of density peaks. To show that this is a generic property of
non-Markovian processes with increments determined by integrals of the
form (Eq.\,\ref{MEM}), we generalize the $[1-e^{-(t-s)/T}]$ term in
Eq.\,(\ref{F2}) by a function $K_T(s,t)$ so that Eq.\,(\ref{EQ}) reads
\begin{equation}\label{G1}
t\,=\,\int_0^tds\,K_T^2(s,t)\,=\,\int_0^\infty
\frac{4\pi k^2\,dk\,P(k)}{(2\pi)^2}\,|W_T(kR)|^2\,.
\end{equation}
Moreover, from the derivation which leads to Eq.\,(\ref{RESOL}) we
found that the resolution variable $t$ is the same for the Brownian
process and our non-Markovian process. The same resolution variable
was also obtained for the Ornstein-Uhlenbeck process (Schuecker et
al. 2001a). We thus generalize Eq.\,(\ref{RESOL}) to a universal
resolution variable
\begin{equation}\label{G2}
t\,=\,\int_0^t\,ds\,=\,\int_0^{1/R}
\frac{4\pi\,k^2\,dk\,P(k)}{(2\pi)^3}\,=\,s[R,P(k)]\,.
\end{equation}
The combination of Eqs.\,(\ref{G1}) and (\ref{G2}) then suggests the
equivalence $K_T(R,P(k))=W_T(kR)$ for $kR\le 1$ and zero
elsewhere. This shows that the kernel of Eq.\,(\ref{F1}) which
determines the increments of the process is closely related to the
profile of the mass filter. 

We further conclude that for merging and accretion processes described
by the comparatively simple integration scheme (Eq.\,\ref{F1}), the
mass filter and thus how much mass is sweeped in from the surrounding
mass by a collapsing region must also depend on the global mass
distribution, and thus on the power spectrum.

Equations (\ref{VOL1}-\ref{RM1}) follow a ``natural'' choice of the
relation between the filter radius, used in the process of halo
detection, and the halo mass as determined by the material which
eventually collapse to form a virialized structure. However, other
choices are also possible (BCEK) and still a matter of debate.

We further note that the application of filters to all spatial points
of a density field, as proposed in all PS-like models, leads to a
large scatter between filter mass and group mass. The introduction of
non-Markovianity follows the same local filtering scheme and is thus
not expected to significantly reduce the scatter on the halo-by-halo
level.

\subsection{Comparison with simulated mass functions}

On the statistical level, PS-like models predict mass functions,
merger rates, formation times, biasing schemes etc.  remarkably
accurately. A comparison of the theoretical mass functions
(Eq.\,\ref{F9}) with the results from large and high-resolution N-body
simulations (e.g., Jenkins et al. 2001) should thus be more fruitful
and should give more important information about the significance of
non-Markovian effects.

The Jenkins et al. (2001) mass function can be used for fluctuation
fields with effective power spectrum slopes between $n_{\rm eff}=-2.5$
and $-1$ at $\sigma=0.5$ in the mass range
$10^{11}-10^{16}\,M_\odot$. Observations suggest a slope of about
$n=-2.0$ (see, e.g., Schuecker et al. 2001b who found $n=-1.8$ on
scales $<100\,h^{-1}\,{\rm Mpc}$ for X-ray clusters of galaxies --
quite consistent with the power spectrum of galaxies, see their
Fig.\,16). We thus compare with our $n=-2$ model.

To be consistent with our ``natural'' choice of the radius-mass
relation (Eq.\,\ref{RM1}), the Jenkins et al. mass function
(e.g. their Eq.\,9) has to be transformed accordingly by using the
$\sigma^2(R(M))$ relation of the non-Markovian process for $n=-2$. To
say it in another way, the non-Markovian effects of the process
(Eq.\,\ref{F1}) can only become apparent after the transformation of
$f(\sigma^2(R))$ to $f(M)$ because non-Markovianity changes the
radius-mass assignment scheme. Therefore, when we want to search for
non-Markovian effects we have to transform $f(\sigma^2)$ of both the
theoretical and the simulated mass function and test different values
of $T$. For this test we multiply the prefactor $16\pi^2$ in the
normalization (Eq.\,\ref{VOL3}) by 1.234 to be consistent with the
standard $\sigma_8$ normalization (obtained with the top-hat filter)
for $T=0$.

The comparison with the Jenkins et al. (2001) mass function within the
mass range $10^{12}-10^{16}\,M_\odot$ (for the lower limit see
Sect.\,\ref{FAIL}) gives the best fit parameter values $T=0.23$ and
$\beta=0.12$ with $\chi^2=1.4$ for three degrees of freedom
(Fig.\,\ref{FIG_M2}). The statistical significance of the
non-Markovian effects strongly depends on the assumed error of the
Jenkins et al. mass function. For errors of 20-30\%, non-Markovian
effects must be regarded as insignificant, whereas for 10\% errors,
non-Markovianity is clearly detected (Fig.\,\ref{FIG_X2}). We thus
conclude that for the most optimistic error estimates of the Jenkins
et al. mass function, its shape suggests the presence of non-Markovian
effects, i.e., effects of the environment on the coherence of the
collapse. Moreover, non-Markovian effects seem to be not very large
and cosmic mass functions with errors better than 10\% are needed for
their clear detection.

\begin{figure}
\vspace{-0.5cm}
\centerline{\hspace{0.0cm}
\psfig{figure=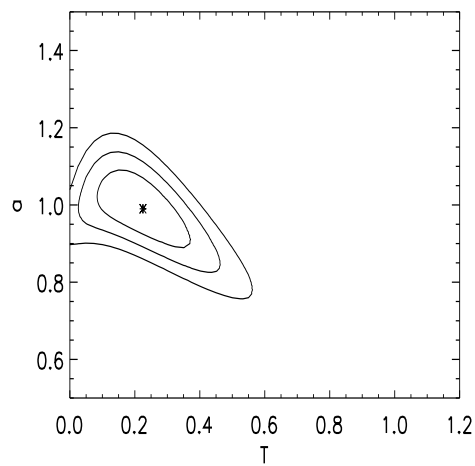,height=7.0cm,width=7.0cm}}
\vspace{-0.5cm}
\caption{\small $\chi^2$ distribution ($1-3\sigma$ contours for three
parameters) in the $\beta=0.12$ plane, assuming 10\% errors of the
Jenkins et al. (2001) mass function.}
\label{FIG_X2D}
\end{figure}

It should be mentioned that we simultaneously tested in
Fig.\,\ref{FIG_X2} for the significance of the scaling parameter $a$
introduced by SMT to account for a failure of the simple counting
argument of the excursion set formalism (R. Sheth, private
communication). The best fit has $a=0.99$ (Fig.\,\ref{FIG_X2D}) which
means that the inclusion of non-Markovian effects does not require
this correction. Therefore, the likelihood contours in
Fig.\,\ref{FIG_X2} are only plotted in the $a=1$ plane.

\subsection{Effects of ellipsoidality}\label{FAIL}

The effects of ellipsoidality on the halo mass function are in general
much larger and are detected for all random errors $\le 30\%$ with
clear significance. The best fit $\beta=0.12$ is lower than
$\beta=0.3$ obtained from simulations (see Sect.\,\ref{FIRST}). In
addition, the model starts to deviate from the Jenkins et al. function
at masses smaller than $10^{12}\,M_\odot$. In order to understand
these failures of the model one should not forget that our goal to
give a full analytic treatment of the problem forces us to approximate
the ellipsoidal collapse by a linear drift. Sheth \& Tormen (2002)
give a simple prescription for approximating the solution to the
general barrier in the Markov case which better fits the low-mass
range. Therefore, a useful direction for future work is to see if
their method works in our particular non-Markovian context also.

Future observed mass functions from e.g. X-ray clusters of galaxies
and lensing studies have high enough precision for detailed studies of
non-Markovian effects expected in the high-mass regime. On the
theoretical side, further studies are in preparation to predict mass
functions also for Cold Dark Matter power spectra to improve the
comparison with observed and simulated mass functions.

\begin{acknowledgements}                                                        
We would like to thank Hans B\"ohringer, Gerard Lemson, Gabriel Pratt,
Ravi Sheth, and the anonymous referee for helpful comments on the
manuscript. GGA is partially supported by INTAS grant No. 00-738 and
PS by DLR grant No.\,50\,OR\,0108.
\end{acknowledgements}

\begin{appendix}

\section{Computation of mass functions}\label{APP}

A summary is given of the equations used for the computation of the
dark matter halo mass functions discussed in the main text. The
meaning of the different symbols are as follows.

$f(\sigma^2)$ dark matter mass function as a function of the variance
$\sigma^2$ of the random mass field or of the diffusion
process. $f(M)$ dark matter halo mass function as a function of mass
$M$. $\delta_{\rm c}$ critical density threshold. $\beta$ parameter
regulating the ellipsoidality of the collapse ($\beta=0$ spherical
collapse). $T$ parameter regulating the coherence of the collapse,
i.e., the deviation from Markovianity ($T=0$ traditional Markovian
excursion set model). $P_0$ amplitude of the scale-invariant power
spectrum in $P(k)=P_0k^n$ with the spectral index $n$. $\bar{\rho}$
mean cosmic matter density. $\Omega_m$ normalized cosmic matter
density. $h$ Hubble constant in units of $100\,{\rm km}\,{\rm
s}^{-1}\,{\rm Mpc}^{-1}$. $t$ pseudo diffusion time or mass resolution
variable. $R$ filter radius. $W(z)$ the principle branch of the
Lambert W-function. $U_1\ldots U_5$, $S_1\ldots S_5$ auxillary
functions, and $U_1'\ldots U_5'$, $S_1'\ldots S5$ their derivatives
with respect to $M$.

The mass functions $f(\sigma^2)$ with the spectral indices $n=0,-1,-2$
have the same form
\begin{equation}
f(\sigma^2)\,=\,\frac{1}{\sqrt{2\pi}}\,\frac{\delta_{\rm
c}}{\sigma^2}\,e^{-\frac{|\delta_{\rm
c}+\beta\sigma^2|^2}{2\sigma^2}}\,,
\end{equation}
and the same $\sigma^2(t)$ dependence
\begin{equation}
\sigma^2\,=\,t\,-\,\frac{3}{2}T\,+\,2Te^{-t/T}\,-\,\frac{T}{2}e^{-2t/T}\,.
\end{equation}
Moreover, the cosmic mass density
\begin{equation}
\bar{\rho}\,=\,2.7755\,\Omega_m\,h^2\,10^{11}\,M_\odot\,{\rm Mpc}^{-3}
\end{equation}
is the same for the power spectra.

\subsection{Mass function for a power spectrum with the spectral index 
$n=0$}\label{N0}
Dark matter halo mass function:
\begin{equation}
f(M)\,=\,f(\sigma^2(R^3(M)))\,\left|\frac{d\sigma^2(R^3)}{dR^3}\right|\,
\left|\frac{dR^3(M)}{dM}\right|\,
\end{equation}
Pseudo diffusion time or mass resolution variable:
\begin{equation}
t\,=\,\frac{P_0}{6\pi^2\,R^3}
\end{equation}
Filter radius-mass relation:
\begin{equation}
R^3\,=\,\frac{M}{6\pi^2\bar{\rho}+\frac{6\pi^2TM}{P_0}\left[1+W(z)\right]}
\end{equation}
Normalization of the matter power spectrum:
\begin{equation}
P_0\,=\,3072\pi^2\left\{\sigma_8^2\,+
\,1.98T\left[1-e^{-(\sigma_8^2/T)^{0.363}}\right]\right\}
\end{equation}
Argument of the Lambert $W$ function.
\begin{equation}
z\,=\,-\exp{\left[-\left(1+\frac{\bar{\rho}P_0}{MT}\right)\right]}
\end{equation}
Gradient of filter radius with mass:
\begin{eqnarray}
\frac{dR^3}{dM}&=&\frac{1}{6\pi^2\bar{\rho}+\frac{6\pi^2TM}{P_0}[1+W(z)]}-\nonumber\\
&&\frac{P_0M\left\{T\left[1+W(z)\right]
+\frac{\bar{\rho}P_0}{eM}\frac{W(z)}{1+W(z)}\right\}}
{6\pi^2\left\{\bar{\rho}P_0+TM\left[1+W(z)\right]\right\}^2}
\end{eqnarray}
Gradient of mass variance with filter radius:
\begin{equation}
\frac{d\sigma^2(R^3)}{dR^3}=-\frac{P_0}{6\pi^2R^6}\left(1-
2e^{-\frac{P_0}{6\pi^2R^3T}}+e^{-\frac{P_0}{3\pi^2R^3T}}\right)
\end{equation}
More information about the Lambert $W$ function can be found in
Corless et al. (1996).

\subsection{Mass function for a power spectrum with the spectral index
$n=-1$}\label{N2}
Dark matter halo mass function:
\begin{equation}
f(M)\,=\,f(\sigma^2(R(M)))\,\left|\frac{d\sigma^2(R)}{dR}\right|\,
\left|\frac{dR(M)}{dM}\right|
\end{equation}
Pseudo diffusion time or mass resolution variable:
\begin{equation}
t\,=\,\frac{P_0}{4\pi^2\,R^2}
\end{equation}
Filter radius-mass relation:
\begin{equation}
R(M)=\frac{(U_1+3\sqrt{U_2})^{1/3}}{U_3}\,+\,
\frac{U_4}{(U_1+3\sqrt{U_2})^{1/3}}\,-\,U_5\,
\end{equation}
Auxillary functions for the computation of radius-mass relation:
\begin{eqnarray}\label{UU}
U_1&=&9\pi^4\bar{\rho}^2P_0^3M-4\pi^6M^3T^3\\
U_2&=&9\pi^8\bar{\rho}^4P_0^6M^2-8\pi^{10}\bar{\rho}^2P_0^3M^4T^3\\
U_3&=&3\cdot 2^{2/3}\pi^2\bar{\rho}P_0\\
U_4&=&\frac{2^{2/3}\pi^2M^2T^2}{3\bar{\rho}P_0}\\
U_5&=&\frac{MT}{3\bar{\rho}P_0}
\end{eqnarray}
Normalization of the matter power spectrum:
\begin{equation}
P_0=256\pi^2\left[\sigma_8^2+1.98\,T\left(1-
e^{-(\frac{\sigma_8^2}{T})^{0.363}}\right)\right]\,{\rm Mpc}^2
\end{equation}
Gradient of filter radius with mass:
\begin{eqnarray}
\frac{dR(M)}{dM}&=&\frac{U_1'+\frac{3}{2}U_2^{-1/2}U_2'}{3U_3(U_1+3\sqrt{U_2})^{2/3}}
+ \frac{U_4'}{(U_1+3\sqrt{U_2})^{1/3}}\nonumber\\
&-&\frac{U_4(U_1'+\frac{3}{2}U_2^{-1/2}U_2')}{3(U_1+3\sqrt{U_2})^{4/3}}\,-\,U_5'
\end{eqnarray}
Derivatives of auxillary functions with respect to mass:
\begin{eqnarray}
U_1'&=&9\pi^4\bar{\rho}^2P_0^3-12\pi^6M^2T^3\\
U_2'&=&18\pi^8\bar{\rho}^4P_0^6M-32\pi^{10}\bar{\rho}^2P_0^3M^3T^3\\
U_4'&=&\frac{2^{5/3}\pi^2MT^2}{3\bar{\rho}P_0}\\
U_5'&=&\frac{T}{3\bar{\rho}P_0}
\end{eqnarray}
Gradient of mass variance with filter radius:
\begin{equation}
\frac{d\sigma^2(R)}{dR}=-\frac{P_0}{2\pi^2R^3}
\left(1-2e^{-\frac{P_0}{4\pi^2R^2T}}+e^{-\frac{P_0}{2\pi^2R^2T}}\right)\\
\end{equation}
The determination of the radius-mass relation neglects the exponential
term in Eq.\,(\ref{VOL6}).

\subsection{Mass function for a power spectrum with the spectral index
$n=-2$}\label{N2}
Dark matter halo mass function:
\begin{equation}
f(M)\,=\,f(\sigma^2(R(M)))\,\left|\frac{d\sigma^2(R)}{dR}\right|\,
\left|\frac{dR(M)}{dM}\right|
\end{equation}
Pseudo diffusion time or mass resolution variable:
\begin{equation}
t\,=\,\frac{P_0}{2\pi^2\,R}
\end{equation}
Filter radius-mass relation:
\begin{equation}
R(M)=\frac{S_1\,S_2^{1/3}\,+\,S_3\,-\,S_4\,S_2^{-1/3}}{S_5}\,
\end{equation}
Auxillary functions for the computation of the radius-mass relation:
\begin{eqnarray}\label{SS}
S_1&=&6(2\pi^4)^{1/3}\\
S_2&=&P_0^3M\,[9\bar{\rho}^2P_0^6+72\pi^4\bar{\rho}P_0^3MT^3+\nonumber\\
&&256\pi^8M^2T^6+3(\bar{\rho}P_0^3+8\pi^4MT^3)\,\times\nonumber\\
&&\sqrt{9\bar{\rho}^2P_0^6+48\pi^4
\bar{\rho}P_0^3MT^3+128\pi^8M^2T^6}\,]\\
S_3&=&48\pi^4P_0MT^2\\ S_4&=&12\cdot
2^{2/3}\pi^{8/3}P_0^2MT(3\bar{\rho}P_0^3+8\pi^4MT^3)\\
S_5&=&36(\pi^2\bar{\rho}P_0^3+8\pi^6MT^3)
\end{eqnarray}
Normalization of the matter power spectrum:
\begin{equation}
P_0=16\pi^2\left[\sigma_8^2+1.98\,T\left(1-
e^{-(\frac{\sigma_8^2}{T})^{0.363}}\right)\right]\,{\rm Mpc}
\end{equation}
Gradient of filter radius with mass:
\begin{eqnarray}
\frac{dR(M)}{dM}&=&\frac{\frac{1}{3}S_1S_2^{-2/3}S_2'+
S_3'-S_4'S_2^{-1/3}+\frac{1}{3}S_4S_2^{-4/3}S_2'}
{S_5}\nonumber\\
&-&\frac{(S_1S_2^{1/3}+S_3-S_4S_2^{-1/3})S_5'}{S_5^2}
\end{eqnarray}
Derivatives of auxillary functions with respect to mass:
\begin{eqnarray}
S_2'&=&9\bar{\rho}^2P_0^9+144\pi^4\bar{\rho}P_0^6T^3M+768\pi^8P_0^3T^6M^2+\nonumber\\
&&(3\bar{\rho}P_0^6+48\pi^4P_0^3T^3M)\times\\
&&\sqrt{9\bar{\rho}^2P_0^6+48\pi^4\bar{\rho}P_0^3MT^3+128\pi^8M^2T^6}+ \nonumber\\
&&\frac{(3\bar{\rho}P_0^6M+24\pi^4P_0^3T^3M^2)
(48\pi^4\bar{\rho}P_0^3T^3+256\pi^8MT^6)}
{2\sqrt{9\bar{\rho}^2P_0^6+48\pi^4\bar{\rho}P_0^3MT^3+128\pi^8M^2T^6}}\nonumber
\end{eqnarray}
\begin{equation}
S_3'=48\pi^4P_0T^2
\end{equation}
\begin{equation}
S_4'=12\cdot2^{2/3}\pi^{8/3}P_0^2T(3\bar{\rho}P_0^3+16\pi^4MT^3)
\end{equation}
\begin{equation}
S_5'=288\pi^6T^3
\end{equation}
Gradient of mass variance with filter radius:
\begin{equation}
\frac{d\sigma^2(R)}{dR}\,=\,-\frac{P_0}{2\pi^2R^2}
\left(1-2e^{-\frac{P_0}{2\pi^2RT}}+e^{-\frac{P_0}{\pi^2RT}}\right)\,.
\end{equation}
The determination of the radius-mass relation neglects the exponential
term in Eq.\,(\ref{VOL2}).

\end{appendix}

\end{document}